\documentclass[twoside,fleqn]{article}
\usepackage{amssymb}
\usepackage{amsbsy}
\usepackage{amsfonts}
\usepackage{epsfig}
\usepackage{espcrc2}

\title{\vspace{-2.0cm}
       {\normalsize DESY 01--170}    \\[-0.2cm]
       {\normalsize October 2001}   \\[0.850cm]
  The strong coupling constant from lattice QCD with $N_f=2$ dynamical
  quarks\thanks{Talk presented by G. Schierholz at Lattice 2001, Berlin, 
  Germany}}  

  \author{S. Booth\address{Edinburgh Parallel Computing Center EPCC, 
   University of Edinburgh, Edinburgh EH9 3JZ, UK},
   M. G\"ockeler\address{Institut f\"ur Theoretische Physik, Universit\"at
   Regensburg, D-93040 Regensburg, Germany},
   R. Horsley\address{NIC/DESY, D-15735 Zeuthen, Germany},
   A.C. Irving\address{Theoretical Physics Division, Department of
   Mathematical Sciences, University of Liverpool, Liverpool \hspace*{0.06cm}
   L69 3BX, UK}, 
   B. Joo\address{Department of Physics, Columbia University, New York, NY
   10027, USA},
   S. Pickles\address{Computer Services for Academic Research CSAR, University
   of Manchester, Manchester M13 9PL, UK}, 
   D. Pleiter$^c$,
   P.E.L. Rakow$^{\rm b}$,\\
   G. Schierholz$^{\rm c,}$\address{Deutsches Elektronen-Synchrotron DESY,
   D-22603 Hamburg, Germany},  
   Z. Sroczynski\address{Fachbereich Physik, Universit\"at Wuppertal, D-42097
   Wuppertal, Germany}, and
   H. St\"uben\address{Konrad-Zuse-Zentrum f\"ur Informationstechnik Berlin,
   D-14195 Berlin, Germany}}

\begin{document}

\begin{abstract}

We compute $\Lambda_{\overline{MS}}$ for two flavors of light dynamical
quarks using non-perturbatively $O(a)$ improved Wilson fermions. We improve 
on a recent calculation by employing Pad\'e-improved two-loop and three-loop 
perturbation theory to convert the lattice numbers to the $\overline{MS}$ 
scheme.
\end{abstract}

\maketitle

\section{THE METHOD}

The running of the strong coupling constant $\alpha_s$ is parameterized by the
$\Lambda$ parameter. In a recent letter~\cite{Booth} we have computed 
$\Lambda_{\overline{MS}}$ from the average plaquette and the scale parameter
$r_0$. An essential step in our calculation was the conversion of the boosted
coupling $g_\square$ to the renormalized coupling $g_{\overline{MS}}$. This 
was done in two-loop tadpole improved perturbation theory, which resulted in an
(estimated) systematic uncertainty of the order of the statistical error. 

We can do better now. In this talk we shall apply Pad\'e techniques to 
approximate an unknown three-loop coefficient in the conversion formula. 
This reduces the systematic uncertainty to an insignificant amount, as we
shall see.  

The calculation of $\Lambda_{\overline{MS}}$ proceeds in three steps:
 
First we compute the average 
plaquette $P= (1/3)\: \langle\mbox{Tr} U_\square\rangle \equiv u_0^4$ from
which we obtain $g_\square$, 
\begin{equation}
\frac{1}{g^2_\square(a)} = \frac{P}{g^2(a)},
\label{gbox}
\end{equation}
where $g(a)$ is the bare lattice coupling. Note that the non-perturbative 
contribution to the plaquette, i.e. the difference between the lattice number
and the perturbative series, is less than $10^{-3}$~\cite{Rakow}. 

In the second step we compute 
$g_{\overline{MS}}$ from $g_\square$. This is the main task. To three loops
we obtain, following the notation of~\cite{Booth},
\begin{eqnarray}
\frac{1}{g_{\overline{MS}}^2(\mu)} \!\!\!\!&=&\!\!\!\!
\frac{1}{g_\square^2(a)} +  2 b_0 \ln a \mu - t^\square_1 \nonumber \\
\!\!\!\!&+&\!\!\!\! ( 2 b_1 \ln a \mu - t^\square_2) \, g^2_\square(a)
\nonumber \\  
\!\!\!\!&+&\!\!\!\! \big( -2 b_0 b_1 \ln^2\!\! a \mu +2(b_2^{\overline{MS}}
+b_1 t^\square_1)\ln a \mu \nonumber \\
\!\!\!\!&-&\!\!\!\! t^\square_3  +p_1t^\square_2\big) \, g^4_\square(a), 
\label{boxcon}
\end{eqnarray} 
where $t_i^\square = t_i - p_i$. We know all coefficients except for 
$t_3^\square$. (See \cite{Booth} for references.) The latter can be computed
from the difference of the four-loop 
coefficients,  $b_3^{\overline{MS}}-b_3^\square$, of the $\beta$ function
\begin{equation}
\beta^{\mathcal S}(g_{\mathcal S})\! =\! -g_{\mathcal S}^3(b_0+b_1g_{\mathcal
  S}^2+b_2^{\mathcal 
  S}g_{\mathcal S}^4 + b_3^{\mathcal S}g_{\mathcal S}^4 +\cdots),
\end{equation}
where ${\mathcal S}=\overline{MS}, \square$ denotes the scheme. We find
\begin{eqnarray}
t_3^\square &=& \big(b_3^{\overline{MS}} - b_3^\square
+2(b_2^{\overline{MS}}t_1-b_2^\square p_1)\nonumber\\
&+&b_1(t_1^2-p_1^2)\big)/2b_0.
\end{eqnarray} 
We know $b_3^{\overline{MS}}$, but we do not know $b_3^\square$. In the 
$\overline{MS}$ scheme the four-loop $\beta$ function turns out to be
very well reproduced by the Pad\'e-improved three-loop $\beta$ function
\begin{equation}
\beta_{[1,1]}^{\mathcal S}(g_{\mathcal S}) = -g_{\mathcal S}^3
\frac{b_0b_1+(b_1^2-b_0b_2^{\mathcal S})g_{\mathcal S}^2}{b_1-b_2^{\mathcal S}
g_{\mathcal S}^2}.
\label{pade}
\end{equation} 
\begin{figure}[tbp]
\vspace*{-0.2cm}
  \begin{center}
    \epsfig{file=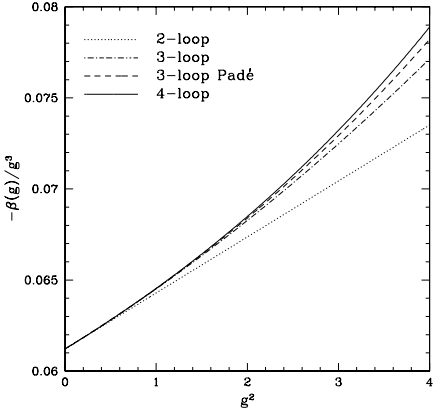,height=6.0cm,width=6.0cm}
\vspace*{-0.75cm}
\caption{The four-loop $\beta$ function for the $\overline{MS}$ scheme
  compared with various approximations.}
    \label{fig1}
  \end{center}
\vspace*{-0.85cm}
\end{figure}
In Fig.~1 we compare (\ref{pade}) with the four-loop expression. We see that
in the region $g_{\overline{MS}}^2 \lesssim 2$, which is the region relevant 
to our
calculation, the Pad\'e-improved three-loop result differs by less than 
$0.02\%$ from the four-loop result. We 
expect that the same is true for the boosted $\beta$ function $\beta^\square$,
which converges at a similar rate as $\beta^{\overline{MS}}$. Thus we may
approximate $\beta^\square$ by $\beta_{[1,1]}^\square$. Expanding (\ref{pade})
gives us then 
\begin{equation}
b_3^\square = \frac{b_2^{\square\, 2}}{b_1}.
\label{b3}
\end{equation} 
We can improve the convergence of the series
(\ref{boxcon}) further by re-expressing it in terms of the tadpole improved
coefficients
\begin{equation}
\widetilde{c}_{SW}= c_{SW} u_0^3,\; a\widetilde{m} = am/u_0.
\end{equation}
This amounts, for example, to replacing every $c_{SW}$ by 
\begin{equation}
\widetilde{c}_{SW}\Big(1+\frac{3}{4}p_1\,g_\square^2
+\frac{3}{4}\big(\widetilde{p}_2
-\frac{1}{6}p_1^2\big)\,g_\square^4\Big). 
\end{equation}
We have already seen in~\cite{Booth} that tadpole improvement reduces 
the coefficients of the two-loop fermionic contribution. We find that tadpole
improvement is equally successful in reducing the coefficients of the 
three-loop contributions.

In the final step we compute $\Lambda_{\overline{MS}}$ from
$g_{\overline{MS}}(\mu)$ using the renormalization group equation
\begin{eqnarray}
\frac{\mu}{\Lambda_{\overline{MS}}}\!\!&=&\!\!\big(b_0
g^2_{\overline{MS}}\big)^{\frac{b_1}{2b_0^2}}
  \exp\Big(\frac{1}{2b_0g^2_{\overline{MS}}}\nonumber\\
\!\!&+&\!\!\int_0^{g_{\overline{MS}}}
  \mbox{d}\xi \big(\frac{1}{\beta^{\overline{MS}}(\xi)}+\frac{1}{b_0\xi^3}
-\frac{b_1}{b_0^2\xi}\big)\Big).
\label{conv}
\end{eqnarray}
For $\beta^{\overline{MS}}$ we use the Pad\'e-improved four-loop
$\beta$ function 
\begin{eqnarray}
\beta_{[1,2]}^{\overline{MS}}\,(g_{\overline{MS}})&=& - g^3_{\overline{MS}}
\Big(b_0\,(b_1^2-b_0b_2^{\overline{MS}})+(b_1^3 \nonumber\\
&-& 2b_0b_1b_2^{\overline{MS}} + b_0^2\,b_3^{\overline{MS}})\,g^2_{\overline{MS}}\Big)\big/
\Big(b_1^2 \nonumber \\
&-& b_0b_2^{\overline{MS}} + (b_0b_3^{\overline{MS}} -
b_1b_2)\,g^2_{\overline{MS}} \nonumber \\
&+& (b_2^{\overline{MS}\, 2}
-b_1b_3^{\overline{MS}})\, g^4_{\overline{MS}}\Big), 
\end{eqnarray}
which is a better approximation than the four-loop $\beta$ function.
The result should be independent of $\mu$ if our approximations of 
$\beta^{\overline{MS}}$ in (\ref{boxcon}) and (\ref{conv}) are consistent.

\section{RESULTS}

We have varied $\mu$ between 4 and 10 GeV and found indeed that
$\Lambda_{\overline{MS}}$ changes by a fraction of a percent only. Finally we
have chosen $a\mu = 2.5$, which corresponds to $\mu \approx 5$ GeV. 
We may estimate the error arising from our approximation (\ref{b3}) 
by replacing $b_3^{\overline{MS}}$ by $b_2^{\overline{MS}\, 2}/b_1$ as well. 
We found that this changes our results by at most 
$0.05\%$.

\begin{table*}[t]
\begin{center}
\vspace{0.2cm}
\begin{tabular}{|c|c|c|c|l|l|l|l|} \hline
$\beta$ & $\kappa_{\mbox{sea}}$ & V  & $c_{SW}$ &\multicolumn{1}{|c|}{$P$} 
&\multicolumn{1}{|c|}{$r_0/a$}& \multicolumn{1}{|c|}{$am$}&
\multicolumn{1}{|c|}{$\Lambda_{\overline{MS}}\,r_0$}\\
\hline
$5.20 $ & $0.1355$ & $16^3 32$ &$2.0171$
 & $0.536294(9)$ & $5.041(40)$ &$0.02364(16)$ &$0.4860(39)$ \\
$5.20 $ & $0.1350$ & $16^3 32$ &$\shortparallel$
 & $0.533676(9)$ & $4.754(40)$ &$0.04586(19)$ &$0.4704(40)$ \\
$5.25 $ & $0.1352$ & $16^3 32$ &$1.9603$
 & $0.541135(24)$& $5.137(49)$ &$0.04268(17)$ &$0.4757(46)$ \\
$5.26 $ & $0.1345$ & $16^3 32$ &$1.9497$
 & $0.539732(9)$ & $4.708(52)$ &$0.07196(20)$ &$0.4430(49)$ \\
$5.29 $ & $0.1355$ & $24^3 48$ &$1.9192$
 & $0.547081(26)$& $5.62(9)$  &$0.03495(12)$  &$0.4917(79)$ \\
$5.29 $ & $0.1350$ & $16^3 32$ &$\shortparallel$
 & $0.545520(29)$& $5.26(7)$  &$0.05348(19)$  &$0.4679(62)$ \\
$5.29 $ & $0.1340$ & $16^3 32$ &$\shortparallel$
 & $0.542410(9)$ & $4.813(45)$ &$ 0.09272(29)$  &$0.4428(42)$\\
\hline
\end{tabular}
\vspace{0.35cm}
\caption{The results for $\Lambda_{\overline{MS}}\,r_0$, together
with $r_0/a$, $P$ and the quark masses $am$, and the parameters of the
simulation.} 
\end{center}
\vspace{-0.75cm}
\end{table*}

To convert our numbers to a physical scale we have used the force parameter
$r_0$. The results are presented in Table 1. The numbers have increased by
$\approx 2\%$ relative to our previous values~\cite{Booth}, which is well 
inside the systematic error that was estimated in that paper. We fit our 
data by
\begin{figure}[b!]
\vspace*{-0.75cm}
  \begin{center}
    \epsfig{file=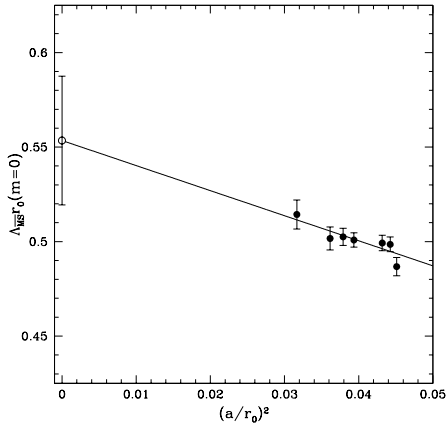,height=6.0cm,width=6.0cm}
    \epsfig{file=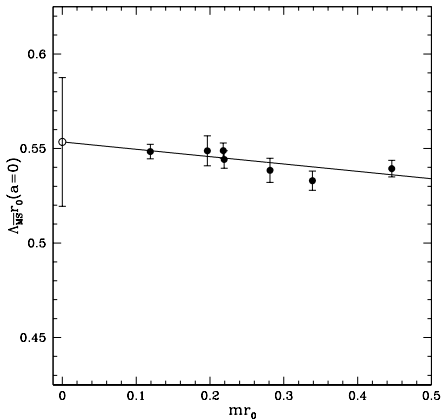,height=6.0cm,width=6.0cm}
\vspace*{-0.85cm}
\caption{$\Lambda_{\overline{MS}}\,r_0 (m=0)$ (top) and 
$\Lambda_{\overline{MS}}\,r_0 (a=0)$ (bottom) together with the fit.}
    \label{fig2}
  \end{center}
\vspace*{-0.75cm}
\end{figure}
\begin{equation}
\Lambda_{\overline{MS}} r_0 =\! A (1 + B am)(1 + C mr_0) + D (a/r_0)^2.
\label{extra}
\end{equation}
(For details of the fit see \cite{Booth}.) In Fig.~2 we show 
$\Lambda_{\overline{MS}}\,r_0 (m=0) = \Lambda_{\overline{MS}}\,r_0 - A(Bam
+Cmr_0+BCam\,mr_0)$ and $\Lambda_{\overline{MS}}\,r_0 (a=0) = 
(\Lambda_{\overline{MS}}\,r_0 - D(a/r_0)^2)/(1+Bam)$ which, if the fit is
successful, should collapse the data points onto a single line each. We see
that this is indeed the case. We also see
that the largest uncertainty arises from the continuum extrapolation, while the
data depend only weakly on the quark mass. In the chiral and continuum limit
our fit gives 
\begin{equation}
\Lambda_{\overline{MS}}r_0=0.553(34) 
\end{equation}
($= A$). Using $r_0=0.5$~fm, we finally obtain 
\begin{equation}
\Lambda_{\overline{MS}}^{N_f=2} = 218(13) \: \mbox{MeV}.
\end{equation}
This is to be compared with our previous result 
$\Lambda_{\overline{MS}}^{N_f=2} = 217(16)(11) \: 
\mbox{MeV}$. 
 
\section{CONCLUSIONS}

Our calculation faces two problems. One is the conversion from the boosted 
to the $\overline{MS}$ scheme. The other is the extrapolation to the chiral 
and continuum limit. We have solved the first problem, and with time and
more data we hope to improve on the extrapolation.


\begin{thebibliography}{9}

\bibitem{Booth} S. Booth et al., Phys. Lett. B {\bf 519} (2001) 229
({\tt hep-lat/0103023}).
\bibitem{Rakow} R. Horsley, P.E.L. Rakow and G. Schierholz, 
{\tt hep-lat/0110210}.

\end{thebibliography}
\end{document}